\documentclass[12pt]{article}
\usepackage{amsmath,amssymb,epsfig}
\renewcommand{\theequation}{\arabic{equation}}
\newcommand{\be}{\begin{equation}}
\newcommand{\ee}{\end{equation}}
\newcommand{\ea}{\end{array}}
\newcommand{\beqa}{\begin{eqnarray}}
\newcommand{\eeqa}{\end{eqnarray}}

\def\CP2{{\mathbb C}P^2}

\def\CDalign#1{\bgroup\vcenter\bgroup\tabskip 2pt 
       \baselineskip 14pt \lineskip 3pt \lineskiplimit 3pt
       \halign\bgroup &\hfill$##$\hfill\crcr
       #1\crcr\egroup\egroup\egroup} 

\renewcommand{\theequation}{\thesection.\arabic{equation}}
\newcommand{\gapproxeq}{\lower .7ex\hbox{$\;\stackrel{\textstyle
>}{\sim}\;$}}
\newcommand{\lapproxeq}{\lower .7ex\hbox{$\;\stackrel{\textstyle
<}{\sim}\;$}}

\newcounter{appendice}
\newcommand{\appendice}
{\setcounter{equation}{0}
\renewcommand{\theequation}{\Alph{appendice}.\arabic{equation}}
\addtocounter{appendice}{1}
{\bf Appendix \Alph{appendice}}
}

\def\thebibliography#1{{\bf REFERENCES\markboth
 {REFERENCES}{REFERENCES}}\list
 {[\arabic{enumi}]}{\settowidth\labelwidth{[#1]}\leftmargin\labelwidth
 \advance\leftmargin\labelsep
 \usecounter{enumi}}
 \def\newblock{\hskip .11em plus .33em minus -.07em}
 \sloppy
 \sfcode`\.=1000\relax}

\begin{document}
\begin{titlepage}
\title{
{\small\hfill FFIA-UV/03-01}\\
NUCLEATION OF \,$^{(4)}R$ BRANE UNIVERSES
\author{
Rub\'en Cordero$^a$ \footnote{Email: cordero@fis.cinvestav.mx} \,
\,and Efra\'\i n Rojas$^b$ \footnote{Email:
efrojas@uv.mx} \\
{\small\it $^a$ Departamento de F\'\i sica, Escuela Superior de F\'\i sica
y Matem\'aticas del I.P.N.}\\
{\small\it Unidad Adolfo L\'opez Mateos, Edificio 9, 07738 M\'exico, D.F., MEXICO}\\
{\small\it $^b$ Facultad de F\'\i sica e Inteligencia Artificial, Universidad
Veracruzana} \\
{\small\it Sebasti\'an Camacho 5; Xalapa, Veracruz; 91000, MEXICO}
} }
\maketitle
\begin{abstract}
The creation of brane universes induced by a totally
antisymmetric tensor living in a fixed background spacetime is
presented, where a term involving the intrinsic curvature of the
brane is considered. A canonical quantum mechanical approach
employing Wheeler-DeWitt equation is done. The probability nucleation
for the brane is calculated taking into account both an instanton
method and a WKB approximation.
Some cosmological implications arose from the model are presented.
\end{abstract}
\end{titlepage}

\section{Introduction}

Nowadays, with the standard cosmology the famous fundamental question,
``where did it all come from?" still it does not have a convincing answer,
reason why a new description is necessary. Cosmologists during long time
have believed that quantum cosmology can shed light on this question
\cite{Vilenkin1, Hawking, Linde, Rubakov} but some
issues are in controversy, e.g.
the lack of an intrinsic time variable in the theory
\cite{Kuchar}, the validity of the minisuperspace
approximation, the problem of cosmological boundary conditions \cite{Hartle}, to
mention something. Among the proposals trying to outline a possible  answer to the
fundamental question, the so-called Brane World Scenaries (BWS)
\cite{Arkani, Randall} became a promising way to understand the birth
and then the evolution of our Universe. Grounded on the proposal that our universe
can be thought as a 4-dimensional spacetime object embedded in an N-dimensional
spacetime, the main physical idea behind of BWS is that the matter
fields are confined to a 3-dimensional space (brane) while the
gravitational fields can extend into a higher-dimensional space (bulk),
where the graviton can travel into the extra dimensions.
Originally proposed to resolve the hierarchy problem,
BWS has been applied to a great
diversity of situations such as dark matter/energy, quientessence,
cosmology, inflation and particle physics.
On other hand, at the formal mathematical level, related
applications of embedding theory such as generation of internal symmetries,
quantum gravity and alternative Kaluza-Klein
theories have been exploited \cite{RT,Maia,Pavsic,Pavsic-Tapia,Rubakov2}.
In the cosmology context there are
predictions of these ideas, that could be tested by astronomical
observations what constitutes one of the several reasons for which it
is so attractive, so that it has predictive power \cite{Supernova}.

In these brane world programs, gravity on the brane can be recovered by
compactifying the extra dimensions \cite{Arkani} or by introducing an AdS
background spacetime \cite{Randall}. However, Dvali, Gabadadze and Porrati
\cite{Dvali1} (DGP) showed that, even in an asymptotically
Minkowski bulk, 4-dimensional gravity can be recovered if one includes a
brane curvature term in the action. Furthermore, DGP considered the $Z_2$
reflection symmetry with respect to the brane getting that gravity, is
4-dimensional on smaller scales than a certain scale, or it is 5-dimensional
on larger distances \cite{Deffayet, Dvali2}.
It is noteworthy that reflection symmetry is not the only possibility
in these models. With regard to the last, several works have been
devoted to antisymmetric cases \cite{Binetruy,Ida,Dolezel,
Gregory,Davis,Carter1,Anchordoqui,Holdom1}, for instance, when the
brane is coupled to a 4-form field \cite{Carter1}. In a pionner work, Brown
and Teitelboim worked out the process of membrane creation by an antisymmetric field motivated
by Schwinger process of pair creation induced for the presence of a electric
field \cite{BT}. Garriga \cite{Garriga2} has also studied the creation of
membranes for this field in a dS background. Others authors have been
interested in brane world creation in AdS spacetime or in other particular
situations \cite{Garriga1,Gorsky1,Odintsov1,Miriam} but, upon our knowledge,
nobody has been devoted to the nucleation of Brane World Universes (BWU)
induced by a 4-form field besides a brane curvature term included in the action.
Generally, BWS are studied mostly for AdS/dS as well as empty (Minkowski)
backgrounds.

In this paper we are going to discuss the nucleation of BWU with a curvature
term induced by a 4-form field in a dS background spacetime. We get
the Friedman like equation when 5-dimensional gravity is fixed and perform
geometric Hamiltonian analysis in order to obtain, by means of canonical
quantization, the corresponding Wheeler-DeWitt equation.
The setup for the induced brane production is
as follows. There is an external homogeneous field that produces a brane;
then, the natural question there, is: what is the probability of such process?
In the present paper we calculate the creation probability for a brane
universe embedded in a de Sitter space,
produced by a 4-form potential gauge field in the same way that the
standard electromagnetic potential bears to a charged particle.
In its quantum analisys we shall use a WKB approximation
attaining the same results by an instanton method. We could try to answer
the question of which one of the universes arose is the more probable
universe produced in this model and if our Universe is one of them, or could
be a very special universe. Parameters of this model must be constrained by
cosmological requirements like nucleosynthesis \cite{Carter1}.

The paper is organized as follows. In Sec. II we present the equations of motion
of a brane with  matter and curvature term that lives in a AdS/dS or Minkowski
bulk when there is no $Z_2$ symmetry and, by means of a limit  equivalent to
the presence of a 4-form field in a fixed background the corresponding equations.
A geometric Hamiltonian approach is done in Sec. III, where the fundamental canonical
structure is obtained and the canonical constraints are listed.
The next step is specialize the general canonical analysis to the case of a
spherical 3-brane floating in an ${\mbox{dS}}_5$ background
spacetime which is the issue of Sec. IV. The last provides the preamble
to obtain the WdW equation in the canonical quantization context, which
is done in Sec. V. The creation probability is calculated in Sec. VI by
two methods, the first is an instanton approach and the other one by means
of a WKB approach for barrier tunneling of the WdW equation. Finally in Sec. VII, we present
our conclusions as well as some perspectives of our work.

\section{The model}
\setcounter{equation}{0}

The effective action that we are interested in the brane world model
corresponds to a 3-brane with a intrinsic curvature term considered
from its worldsheet and no $Z_2$ symmetry in the
presence of a fixed background spacetime.
We consider the following action
\begin{equation}
S = \int d^5 y \sqrt{-g} \left( \frac{1}{2k}{} ^{(5)}{\cal R} + {\cal L}_m \right) +
\int d^4 x \sqrt{-\gamma} \left( \frac{1}{2k'} {\cal R} - L_m \right)
\end{equation}
where ${\cal L}_m$ and $L_m= \rho_v$ stand for matter Lagrangians for the bulk
and the brane, respectively. In our case, we will consider those as cosmological
constants. The constants $k=M^{2-N} _{(N)}$ and $k' = M^{-2} _{(4)}$, where $M_{(4)}$ and
$M_{(N)}$ are the brane Plank and bulk masses. $N$ denotes the dimension of the
bulk. The respective equations of motion
for the brane are \cite{Ida},
\begin{eqnarray}
[K]\gamma_{a  b} - [K_{ab}]&=& k{\cal T}_{ab}, \\
{\widetilde{T}}^{ab} <K_{a b}> &=& [{\cal T}_{nn}], \\
\nabla_{a} ( T^a {}_b ) &=& -[\widetilde{T}_{b n}] .
\end{eqnarray}
where $K_{ab}$ is the extrinsic curvature of the brane, $\gamma_{ab}$
denotes the worldsheet metric.
${\cal T}_{ab} = ({\cal T}_{bulk})_{\mu \nu} e^\mu {}_a e^\nu {}_b$,
${\cal T}_{a n} = ({\cal T}_{bulk})_{\mu \nu} e^\mu{}_a n^\nu$   and
${\cal T}_{nn} = ({\cal T}_{bulk})_{\mu \nu}n^\mu n^\nu$ are the
projections onto the worldsheet of the bulk energy-momentum tensor.
The square and angular brackets
represent the difference and the
average of the corresponding embraced
quantity, on the two sides of the brane, respectively, i.e.,
$[K_{ab}] = K^+ _{ab} - K^- _{ab}$ and $<K_{ab}> =
\frac{1}{2}(K^+ _{ab} + K^- _{ab})$, where
`+' and `-' denote the exterior and interior of the brane.

Taking into account that the bulk energy momentum tensor has the form
\begin{equation}
{\cal T}^{\pm} _{\mu \nu} = - k^{-1} \Lambda^{\pm}g_{\mu \nu} ,
\end{equation}
and by means of the generalized Birkhof theorem, the 5-dimensional FRW metric
can be written as
\begin{equation}
dS^2 _5 = -A_{\pm} d\tau^2 + A^{-1} _{\pm}da^2 + a^2d\Omega^2 _3\,,
\label{eq:dS5}
\end{equation}
where
\begin{equation}
A_{\pm} = \kappa - \frac{\Lambda^{\pm}}{6}a^2 -
\frac{2{\cal M}^{\pm}}{M^3 _{(5)}a^2}\,,
\label{eq:A}
\end{equation}
and $d\Omega^2 _3$ denotes the metric of a 3-sphere, $a$ is the
cosmic scale factor and ${\cal M}^\pm$ is the mass. Furthermore,
in the cosmic time gauge the
4-dimensional metric on the brane reduces to
\begin{equation}
dS^2 _4 = - dt^2 + a^2d\Omega^2 _3.
\end{equation}
Using the junction conditions, and
due to we have isotropy and homogeneity in (\ref{eq:dS5}),
matter can be parametrized completely via a perfect fluid brane
energy-momentum tensor
\begin{equation}
T^a {}_b = {\mbox{diag}}(-\rho, P, P, P),
\end{equation}
so the relevant equations of motion for the model are the following
\begin{eqnarray}
\left( {\dot{a}}^2 + A_{-} \right)^{1/2} - \left( {\dot{a}}^2
+ A_{+} \right)^{1/2}
&=& \frac{ka}{3}\left( \rho - \frac{3({\dot a}^2 + 1)}{k'a^2} \right) , \\
\dot{\rho} + 3\frac{\dot a}{a}(\rho + P)&=& 0 .
\label{eq:motion}
\end{eqnarray}
Last equation represents the energy-momentum conservation on the brane.
The former system was discussed in \cite{Ruben-Alex} where several interesting cases
were treated. Suppose now ${\cal M}^- = 0$, $\rho = {\mbox{const}}$, and
consider at the same time, the limits of fixed bulk gravity, $M_{(5)}
\rightarrow  \infty $ and, $\Lambda^+ \rightarrow \Lambda^-$ but satisfying the
following relation
\begin{equation}
{\mbox{\large{Lim}}}_{\tiny{{(M_{(5)},  \Lambda^+)} \rightarrow {(\infty, \Lambda^-)}}}
(\Lambda^+ -\Lambda^-)M^3_{(5)} = \alpha  ,
\end{equation}
so, expanding the second term of the LHS of Eq. (\ref{eq:motion}), this
equation transform to
\begin{equation}
\left(\frac{\rho}{3} - M^2 _{(4)}\frac{\dot{a} +1}{a^2}\right)
\left(\frac{\dot{a} +1}{a^2}- \frac{\Lambda}{6}
\right)^{1/2} = \frac{\alpha}{12} + \frac{{\cal M}}{a^4}.
\label{eq:energy}
\end{equation}
In order to get the Friedman like
equation we define a $\Upsilon$ quantity through its definition
\begin{equation}
\frac{\dot{a} +1}{a^2} \equiv \frac{\rho}{3M^2 _{(4)}}\Upsilon \equiv H^2\Upsilon .
\end{equation}
Note that $\Upsilon$ is only a function of $a$ and it is a solution of the following
relation
\begin{equation}
M^4 _{(4)}(1 -\Upsilon)^2\left( \Upsilon - \frac{\Lambda}{6H^2} \right) =
H^{-6} \left(\frac{\alpha}{12} +
\frac{{\cal{M}}}{a^4}\right)^2 .
\label{eq:first}
\end{equation}
As we will see below, this approach is equivalent to a brane
interacting with a 4-form field and propagating in a fixed
background spacetime.

\section{Hamiltonian Approach}
\setcounter{equation}{0}

The Hamiltonian framework has been a fundamental prop in the
study of the dynamics of field theories besides of appoint oneself
a preliminary step towards canonical quantization in physical
theories. Knowingly of previous fact, canonical quantization is the
oldest and most conservative approach to quantization which we would like to
develop in order to attain the quantum cosmology emerged from our BWU model.
To carry out the previous thing, we must begin by
casting the theory in a canonical fashion, then we shall
proceed to its quantization.

To begin with, we are going to mimic the well known ADM
procedure for canonical gravity to get a hamiltonian description
of the brane. We shall assume that the worldsheet $m$ admits a foliation,
i.e.,  we will begin with a time like 4-manifold $m$ topologically
$\Sigma \times R$, equipped with a metric $\gamma_{ab}$, such
that $m$ is an outcome of the evolution of a space like 3-manifold
$\Sigma_t$, representing ``instants of time", each of which is
diffeomorphic to $\Sigma$. Then we shall proced to identify the several
geometric quantities inherent to the hypersurface $\Sigma_t$. The
ADM decomposition of the action, computation of the momenta as well as
the recognition of the constraints are the succesive stages.

\subsection{Model ADM decomposed}

Leaning in results achieved in \cite{Supercond, Sardinia, Chiral},
we are going to display the standard
procedure. We start considering the action
\begin{equation}
S = \frac{k_1}{2} \int_m \sqrt{-\gamma} \,({\cal R} + \Lambda_b) +
\frac{k_2}{4!} \int_m \sqrt{-\gamma} A_{\mu \nu \rho \sigma }
\epsilon^{\mu \nu \rho \sigma } \,,
\label{eq:RMaction}
\end{equation}
where ${\cal R}$ is the Ricci scalar curvature of the
worldsheet $m$, $k_1 = M_{(4)} ^2$ and $\Lambda_b = - 2\rho_v/M_{(4)} ^2$
being the cosmological constant
on the brane. $A_{\mu \nu \rho \sigma }$ is a gauge 4-form
Ramond-Ramond field onto the bulk, $\mu ,\nu = 0,1,\dots,N-1$.
$\epsilon^{\mu \nu \rho \sigma }$ is an antisymmetric bulk tensor
which can be expressed in terms of
the worldsheet Levi-Civita tensor as $\epsilon^{\mu \nu \rho \sigma }=
\epsilon^{abcd}e^\mu{}_a e^\nu{}_b e^\rho {}_c e^\sigma {}_d$, where
$e^\mu {}_a$ denotes the tangent vectors to the worldsheet, $a,b= 0,1,2,3$.
$k_2$ is the coupling constant between the brane and the
antisymmetric tensor.

Before going on, we would like to glimpse onto the ADM
decomposition of some important geometric quantities defined onto
the branes in our geometrical approach.
In the Appendix we have included notation and some important
facts for embedding theories to have reference of the material
useful through the paper.

Taking into account the Gauss-Codazzi relations for the
embedding of $\Sigma_t$ in $m$, Eqs. (\ref{eq:GW4}) and
(\ref{eq:GW5}), up to a divergence term we have an equation involving
the curvatures either extrinsic and intrinsic
\begin{equation}
{\cal R} = R + ( k_{AB}k^{AB} - k^2)\,,
\label{eq:GC}
\end{equation}
where $R$ denotes the intrinsic curvature\footnote{We will adhere to Wald's
convention concerning the definitions of Riemannian curvature, namely,
$2 \nabla_{[a}\nabla_{b]}t^c = - {\cal R}_{abd}{}^c \,t^d$ \,\cite{Wald}}
of $\Sigma_t$ which does not have any dependence of the velocity and
$k_{AB}$ its extrinsic curvature associated with the unit timelike normal
$\eta^\mu$, given by
\begin{eqnarray}
k_{AB} &=& - g_{\mu \nu}\eta^\mu ({\cal D}_A \epsilon^\nu {}_B
+ \Gamma^\mu _{\alpha \beta} \epsilon^\alpha {}_A \epsilon^\beta {}_B)
\nonumber \\
       &:=& - g_{\mu \nu} \eta^\mu \tilde{{\cal D}}_A \epsilon^\nu {}_B \,.
\label{eq:ktensor}
\end{eqnarray}
Besides of (\ref{eq:ktensor}), in $\Sigma_t$ we have another curvature
tensor associated with the $i$th unit normal $n^{\mu \,i}$
\begin{equation}
K_{AB} ^i = -  g_{\mu \nu} n^{\mu\,i} \tilde{{\cal D}}_A \epsilon^\nu {}_B \,,
\label{eq:Ktensor}
\end{equation}
where $g_{\mu \nu}$ denotes the background spacetime metric and
$i=1,2,\dots , N - d; \,A,B=1,2,3$.
Note that the configuration space consists
of the embedding functions $X^\mu$ for the brane, instead of 3-metrics
as is customary in the ADM approach for general relativity.

In order to simplify the computations below, the next relations
will be more useful since the velocities appear explicitly
\begin{eqnarray}
\kappa_{AB}&=& N\,k_{AB} \\
&=& - g_{\mu \nu}\dot{X}^\mu \tilde{{\cal D}}_A \epsilon^\nu {}_B
\,. \nonumber
\end{eqnarray}
For canonical purposes will be useful the next time
derivative
\begin{equation}
\frac{\partial N}{\partial \dot{X}^\mu} = -
\eta_\mu = - \,g_{\mu \nu} \eta^\nu\,.
\end{equation}
As before, we will need the derivatives of the extrinsic
curvature
\begin{eqnarray}
\frac{\partial \kappa_{AB}}{\partial \dot{X}^\mu} &=&
- g_{\mu \nu} \tilde{{\cal D}}_A \epsilon^\nu {}_B\\
&=& - k_{AB} \,\eta_\mu + K_{AB} ^i \,n_{\mu \,i}, \nonumber
\end{eqnarray}
where in the second line on the RHS we have used the
Gauss-Weingarten equations (\ref{eq:GW1}).

The ADM decomposed action (\ref{eq:RMaction}) now looks like
\begin{equation}
S= \int_{\Sigma_t} \int_R \frac{k_1}{2} N \sqrt{h} \left[ \bar{R} + k_{AB} k^{AB} - k^2
\right]
+
 \int_{\Sigma_t} \int_R \frac{k_2}{3!}\, A_{\mu \nu \rho \sigma }
\dot{X}^\mu \epsilon^\nu{}_A
\epsilon^\rho{}_B \epsilon^\sigma{}_C \, \varepsilon^{ABC}
\end{equation}
where we have defined $\bar{R} := R + \Lambda_b$ and $h$ is the determinant of the
hypersurface metric $h_{AB}$ and $\varepsilon^{ABC}$ is the $\Sigma_t$ Levi-Civita
antisymmetric symbol.

\subsection{Primordial tensor}

We define for convenience the following symmetric
tensor which is independent of the velocities
\begin{eqnarray}
\Theta^{\mu}{}_ \nu &:=& \,(h^{AB}h^{CD} - h^{AC}h^{BD})\,
\tilde{{\cal D}}_A \epsilon^\mu {}_B
\tilde{{\cal D}}_C \epsilon_{\nu \,D}
\nonumber
\label{eq:tensor}
\\
&=& (k^2 - k_{AB}k^{AB})\,\eta^\mu \eta_\nu
- (kL^i - K_{AB} ^i k^{AB})\,n^\mu {}_i \eta_\nu \nonumber \\
&-& (kL^i - K_{AB} ^i k^{AB})\,\eta^\mu n_{\nu \,i}
+  (L^i L^j - K_{AB} ^i K^{AB\,j})\, n^\mu{}_i n_{\nu \,j} ,
\end{eqnarray}
where $L^i$ denotes the trace of the curvature $K_{AB} ^i$, i.e.,
$L^i = h^{AB} K_{AB} ^i$. This tensor will keep track of the dynamics of the
theory as we will below. The tensor (\ref{eq:tensor}) was
previously defined in \cite{Karasik-Davidson} where a Hamiltonian
analysis for geodetic brane gravity was performed.
We will have in mind some ideas of the classical approach
developed there.

Some of the important properties we are interested from the tensor
(\ref{eq:tensor}) are the following
\begin{eqnarray*}
\Theta^{\mu}{}_\alpha \epsilon^\alpha {}_A &=& 0 \,, \\
\Theta^{\mu}{}_\alpha \dot{X}^\alpha &=& -  N
(k^2 - k_{AB} k^{AB})\,\eta^\mu +  N (k L^i - K_{AB} ^i k^{AB} )
\,n^\mu {}_i   ,     \\
g_{\mu \nu}  \dot{X}^\mu
\Theta^{\nu}{}_\alpha \dot{X}^\alpha &=& N^2 (k^2 -
k_{AB} k^{AB} )  .
\end{eqnarray*}
We shall adopt the notation $\dot{X}\cdot \Theta \cdot \dot{X}:=
g_{\mu \nu}  \dot{X}^\mu \Theta^{\nu}{}_\alpha \dot{X}^\alpha$
throughout the paper. Taking advantage of the previous results
we are able to rewrite the Lagrangian density as follows
\begin{equation}
{\cal L} = \frac{k_1}{2} N \sqrt{h} \left[  \bar{R}  - \frac{1}{N^2}\,
\dot{X}\cdot \Theta \cdot \dot{X}
\right]   +
\frac{k_2}{3!}A_{\mu \nu \rho \sigma}
\dot{X}^\mu \epsilon^\nu {}_A  \epsilon^\rho {}_B
\epsilon^\sigma {}_C \, \varepsilon^{ABC}.
\end{equation}

Using the tensor (\ref{eq:tensor}),
the momenta associated to the embedding functions are the following
\begin{eqnarray}
P_\mu &=& \frac{\partial {\cal L}}{\partial \dot{X}^\mu} \nonumber \\
&=& - \frac{k_1}{2} \sqrt{h}  \,
\left\{  \left[ \bar{R} + \frac{1}{N^2}\,\dot{X} \cdot \Theta
\cdot \dot{X}  \right]\, \eta_\mu
+ \frac{2}{N} \Theta_{\mu \nu} \dot{X}^\nu \right\}
+ \frac{k_2}{3!} \,A_{\mu \alpha \beta \gamma}\,
\bar{\varepsilon}^{\,\alpha \beta \gamma}  ,
\label{eq:P}
\end{eqnarray}
where we have defined the $\Sigma_t$-antisymmetric tangent tensor
$\bar{\varepsilon}^{\mu \nu \rho}= \varepsilon^{ABC}\epsilon^\mu{}_A
\epsilon^\nu{}_B \epsilon^\rho{}_C$ with normalization
$\bar{\varepsilon}^{\mu \nu \rho} \bar{\varepsilon}_{\mu \nu \rho}= 3!$ .

\subsection{Canonical Constraints}

Due to we have in hands an invariant reparametrization theory,
a natural question to ask is what its inherited primary constraints are.
This is part of the chore for constrained field theories.
According to the standard Dirac-Bergmann algorithm, we will get the constraints
from the momenta (\ref{eq:P}).  It is convenient for the computation,
define the matrix $\Psi^{\mu}{}_\nu := \Theta^{\mu}{}_\nu - \lambda
g^{\mu}{}_\nu$ where $\lambda (x)$ is a not dynamical
field which is gauge dependent \cite{Karasik-Davidson}, to be found.
If we assume that the form of momenta have the following pattern,
\begin{equation}
P_\mu = -  \sqrt{h} k_1 \left( \Theta - \lambda \,g\right)_{\mu \nu}
\,\eta^\nu + \frac{k_2}{3!} \,A_{\mu \alpha \beta \gamma}\,
\bar{\varepsilon}^{\,\alpha \beta \gamma}\,,
\label{eq:P3}
\end{equation}
we are free to compare both expressions (\ref{eq:P}) and (\ref{eq:P3})
to get a condition to be satisfied
\begin{equation}
  \bar{R} + \eta \cdot \Theta \cdot \eta + 2 \lambda = 0 \,.
\label{eq:relation}
\end{equation}
This expression will metamorphose in a primary constraint after we
express it in terms of phase space variables.

Profitable is the introduction of the field $\lambda(x)$ since we can solve
Eq.(\ref{eq:P3}) for the timelike unit normal vector
\begin{equation}
\eta^\mu = \frac{-1}{\sqrt{h} k_1}\,\left( \Psi ^{-1}
\right)^{\mu}{}_\alpha g^{\alpha \beta} {\cal P}_\beta \,,
\end{equation}
where we have defined ${\cal P}_\mu = P_\mu - \frac{k_2}{3!}\,
A_{\mu \alpha \beta \gamma}\,\bar{\varepsilon}^{\,\alpha \beta \gamma} $,
but we have to pay a price which is enlarge the number of
constraints as we will see below. Inserting this form of the unit
time-like vector in the relation (\ref{eq:relation}), we get the
main scalar primary constraint. In a similar way, inserting
$\eta^\mu$ in its square relation, $g(\eta,\eta)=-1$, we have
another scalar constraint.

The complete set of primary constraints we have in hand
are the following
\begin{eqnarray}
C_0 &=& {\cal P} \cdot (\Psi^{-1}) \cdot {\cal P} +
h\lambda_0 k_1 ^2 = 0 \,,
\label{eq:C0}
\\
{\cal C}_0 &=& {\cal P} \cdot (\Psi^{-2}) \cdot {\cal P}  +
h k_1 ^2 = 0 \,,
\label{eq:C00}
\\
C_A &=& {\cal P}_\mu \epsilon^\mu {}_A = 0 \,,
\label{eq:CA} \\
{\cal C}_\lambda &=& P_\lambda = 0\,,
\label{eq:Clambda}
\end{eqnarray}
where we have defined $\lambda_0 = \lambda + \bar{R}$.
The third constraint is the always inherited constraint to the parametrized
theories while the last one came from the fact that $\lambda$ is not a
dynamical field, i.e., its time derivative does not
appear in the Lagrangian. It is worthy mention that the constraint
${\cal C}_0$ is a byproduct of $C_0$
taking advantage of the identity
$\partial (\Psi^{-1})^\mu {}_\nu /\partial \lambda
= (\Psi^{-2})^\mu {}_\nu \,.
$


\section{Brane Universe Floating in a de Sitter Space}
\setcounter{equation}{0}

The main idea in this section is adapt the previous dynamical
description to the case of a spherical brane immersed in a
specific background spacetime in order to apply the quantum
approach to our BWS model.

Consider a 3-dimensional spherical brane evolving in a de
Sitter 5-dimensional background spacetime, $dS_5 ^2 = - A_\pm
\,d\tau^2 + A_\pm ^{-1}\,da^2 + a^2 d\Omega_3 ^2 $, where
$A_\pm$ is given by (\ref{eq:A}). The worldsheet generated by the
motion of the brane can be described by the following
embedding
\begin{equation}
x^\mu = X^\mu (\tau, \chi, \theta, \phi) =
\left(
\begin{array}{c}
t(\tau) \\
a(\tau) \\
\chi \\
\theta \\
\phi
\end{array}
\right)  \,.
\label{eq:embedding}
\end{equation}
The line element induced on the worldsheet is given by
\begin{equation}
ds^2 = (- A_\pm \dot{t}^2 +  A_\pm ^{-1}\dot{a^2}) \,d\tau^2 + a^2 \,d\chi^2 +
a^2 \sin^2 \chi \, d\theta^2 + a^2 \sin^2 \chi \sin^2 \theta\, d\phi^2 ,
\label{eq:metric}
\end{equation}
where the dot stands for derivative with respect to cosmic time $\tau$.
For convenience in notation  we define $\Delta = -A_{\pm} \dot{t}^2 +
A_{\pm} ^{-1}\dot{a}^2$. The frecuently
appealed cosmic gauge will be set up by $\Delta = -1$.

In order to evaluate the extrinsic curvature tensors involved in our
approach, (\ref{eq:ktensor}) and (\ref{eq:Ktensor}), we need the
orthonormal $\Sigma_t$ basis
$$
\eta^\mu = \frac{1}{\sqrt{-\Delta}}\left( \dot{t}, \dot{a}, 0 , 0 , 0 \right)\,,
\quad \quad n^\mu = \frac{1}{\sqrt{-\Delta}}\left( A_{\pm} ^{-1}
\,\dot{a}, A_{\pm} \,\dot{t}, 0 , 0 , 0 \right)\,.
$$

The only nonvanishing components for the extrinsic curvatures are
\begin{eqnarray*}
k_{\chi \chi}&=& \frac{a\dot{a}}{(-\Delta)^{1/2}}
\quad \quad \quad \quad \quad \quad
K_{\chi \chi}= \frac{a\dot{t}}{(-\Delta)^{1/2}}A_{\pm} \\
k_{\theta \theta}&=& \frac{a\dot{a}}{(-\Delta)^{1/2}} \sin^2 \chi
\quad \quad \quad
K_{\theta \theta}= \frac{a\dot{t}}{(-\Delta)^{1/2}}A_{\pm}\sin^2 \chi  \\
k_{\phi \phi}&=& \frac{a\dot{a}}{(-\Delta)^{1/2}} \sin^2 \chi \sin^2
\theta
\quad
K_{\phi \phi}= \frac{a\dot{t}}{(-\Delta)^{1/2}}A_{\pm}\sin^2 \chi
\sin^2 \theta  \,.
\end{eqnarray*}

It is a straightforward task compute the tensor (\ref{eq:tensor})
for the present case, which give us
\begin{equation}
\left(\Theta\right)^{\mu}{}_\nu=
\left(
\begin{array}{ccc}
0 & 0 & 0 \\
0 & \frac{6}{a^2}\,A_{\pm} & 0 \\
0 & 0 & 0_{3\times 3}
\end{array}
\right)_{5\times 5}\,.
\label{eq:tensor-matrix}
\end{equation}
The next task is compute the matrix $\Psi$ so, in order to know $\Psi$
is necessary evaluate $\lambda$. It is easily calculated from the relation
(\ref{eq:relation}), given by
\begin{equation}
\lambda = - \frac{1}{2a^2}\left( 6 + \Lambda_b a^2 +
\frac{6\dot{a}^2}{(-\Delta )}\right).
\label{eq:lambda}
\end{equation}
This seems contradict the functional dependence for the field previously assumed,
but we are free to implement an artistry to convert the velocity dependence to
the right form by means of the generalized evolution equation,
$({\dot{a}}^2 + 1)/a^2 = \Upsilon H^2$, avoiding any misunderstanding.

We turn now to compute a first integral for our specific model. This is
performed from (\ref{eq:P}) by setting up $P_0$ proportional
to the brane energy, $P_0 := 3 E \Phi = 3 E (\sin^2 \chi
\sin \theta)$. Furthermore, since we have a homogeneous isotropic space
in (\ref{eq:metric}), we can invoke the typical value
$A_{0\chi \theta \phi} = \frac{F}{4} a^4 \Phi$
for the gauge field, which is supported by some kind of cosmological solutions
\cite{Carter1, Kiritsis}, where F is a constant and the corresponding
gauge independent field tensor
$F_{\mu \nu \rho \delta \gamma} = 5 \nabla_{[\mu}A_{\nu \rho \delta \gamma]}$
is expressed in terms of it
$F_{\mu \nu \rho \delta \gamma} = F \epsilon_{\mu \nu \rho \delta \gamma}$.
Explicitly, we have
\begin{equation}
P_0 = \frac{3 k_1 a\dot{t} \Phi A_{\pm}}{\sqrt{-\Delta}} \,
\left( 1 + \frac{\Lambda_b}{6}a^2 + \frac{{\dot{a}}^2}{(-\Delta)} \right)
+ \frac{k_2 F}{4} a^4\Phi \,.
\end{equation}
Now, taking into account the generalized evolution equation and
$\Lambda_b$ being the cosmological constant
on the brane, we find the desired result
\begin{equation}
E = M^2 _{(4)}a^4  H^3 \left( \Upsilon  - \frac{\Lambda}{6H^2}
\right)^{1/2}\left(  \Upsilon - 1 \right) + \frac{k_2 F}{12} \,a^4 \,,
\label{eq:energyagain}
\end{equation}
where $\Lambda$ is the cosmological constant living in the bulk appearing
in Eq. (\ref{eq:A}) and we have used the cosmic gauge in the last step.
Note that (\ref{eq:energyagain}) is in keep with Eq. (\ref{eq:first}),
confirming equivalence with the limit process developed in Sect. 2.


\section{Wheeler-DeWitt equation}
\setcounter{equation}{0}

We turn now in this section to develop the quantum description for
our specific problem.
The canonical quantization procedure is well known so, just remain
apply the recipe in the matter of our case.

We shall set $P_\mu \rightarrow -i\frac{\delta}
{\delta X^\mu}$ in such a way that scalar constraints
(\ref{eq:C0}) and (\ref{eq:C00}) transform into quantum equations
\begin{eqnarray}
\left( -i\frac{\delta}{\delta X^\mu} -  p_{A\,\,\mu}\right)(\Psi ^{-1})^{\mu \nu}
\left( -i\frac{\delta}{\delta X^\mu} - p_{A\,\,\mu}\right)\psi &=& - h \lambda_0 k_1 ^2
\,\psi \,,
\label{eq:0quantum} \\
\left( -i\frac{\delta}{\delta X^\mu} - p_{A\,\,\mu}\right)(\Psi ^{-2})^{\mu \nu}
\left( -i\frac{\delta}{\delta X^\nu} - p_{A\,\,\nu}\right)\psi &=& -  h k_1 ^2
\,\psi \,,
\label{eq:quantum}
\end{eqnarray}
where we have defined $p_{A\,\,\mu} :=
k_2 \,A_{\mu \alpha \beta \gamma} \bar{\varepsilon}^{\alpha \beta \gamma}/3!$\,.

Specializing to the embedding (\ref{eq:embedding}) and
having in mind the matrix (\ref{eq:Psicosmicgauge}) in the cosmic
gauge, we are able to get the inverse matrix
\begin{equation}
(\Psi^{-1})^{\mu}{}_\nu \equiv \left(
\begin{array}{ccc}
A&
0 & 0\\
0 & B & 0 \\
0 & 0 & N_{3 \times 3} ^{-1}
\end{array}
\right) =
\left(
\begin{smallmatrix}
\frac{-1}{3H^2(1 - \Upsilon )} & 0 & 0 \\
0 & \frac{a^{2}}{3  [- H^2 a^2 (1 - \Upsilon ) + 2 A_{\pm}]} & 0  \\
0 & 0  & N_{3 \times 3} ^{-1}
\end{smallmatrix}
\right) ,
\end{equation}
in such a way that (\ref{eq:0quantum}) and (\ref{eq:quantum})
transform in the pair of relations
\begin{eqnarray}
- A_{\pm} ^{-1} A{\widetilde P}^2 _0 \,\psi + A_{\pm} B{\widetilde P}^2 _1 \,
\psi &=& -h \lambda_0 k_1^2
\,\psi ,\\
- A_{\pm} ^{-1} A^2{\widetilde P}^2 _0 \,\psi + A_{\pm} B^2{\widetilde P}^2 _1 \,
\psi &=& - h k_1 ^2 \, \psi  ,
\end{eqnarray}
where we introduce the notation ${\widetilde P}_\mu = -i
\frac{\delta}{\delta X^\mu} - p_{A\,\,\mu}$.
Taking into account the value $\lambda_0 = 3 \left[ - H^2 (1 + \Upsilon) + \frac{2}{a^2}
\right]$ expressed in the cosmic gauge, the couple of quantum relations can be rewritten
as,
\begin{eqnarray}
{\widetilde P}^2 _0 \,\psi &=& k^2 _1 (3\Phi)^2 a^8 H^6 ( 1 -\Upsilon)^2 \left( \Upsilon -
\frac{\Lambda}{6H^2}\right)\, \psi ,    \\
{\widetilde P}^2 _1 \,\psi &=& -k_1 ^2 (3\Phi)^2 a^2 \frac{(1 - H^2 \Upsilon a^2)
[H^2a^2(1 - \Upsilon) -2 + \frac{\Lambda a^2}{3}]^2}{(1 - \frac{\Lambda a^2}{6})^2}\,\psi \,.
\end{eqnarray}
At this time, we are more interested in identify the potential governing the dynamics
of our model instead of solve exactly the WdW equation so, to get insight
we propose the wave function of separable form, $\psi(t,a) = \psi_1(t)\Psi(a)$.
The WdW equation adquires the form
\begin{equation}
-\frac{\partial^2 \Psi}{\partial a^2} = \frac{a^2 M_{(4)} ^4 \left[ 2 -\frac{\Lambda a^2}{3}
+ \left( \Upsilon -1 \right) H^2a^2\right]^2 \left( -1 + \Upsilon H^2a^2 \right) }
{\left( 1-\frac{\Lambda a^2}{6}\right)^2}\Psi ,
\label{eq:WdW}
\end{equation}
accompanied by the energy equation
\begin{equation}
\left(E - \frac{k_2 F}{12}  a^4\right)^2 = H^6a^8 M_{(4)} ^4 (1 -\Upsilon)^2\left(\Upsilon -
\frac{\Lambda}{6H^2}\right)  ,
\label{eq:moreenergy}
\end{equation}
where, as before, we have assumed  ${\widetilde P}_0 = (3 \Phi) E$.

\section{Nucleation Rate}
\setcounter{equation}{0}

At this stage, we are ready to compute the creation probability
which the universe could be created. Some simplifications are necessary
due to the general problem itself is hard to solve.

From WdW equation (\ref{eq:WdW}), is easily read off the potential which
is subjected the model (\ref{eq:RMaction})
\begin{equation}
V(a) = \frac{a^2 M_{(4)} ^4 [2 - \frac{\Lambda a^2}{3} + (\Upsilon -1)H^2 a^2]^2
(1 - \Upsilon H^2a^2)}{(1 - \frac{\Lambda a^2}{6})^2} \,.
\label{eq:potential}
\end{equation}
Note that this is a very hard expression to work out if one is interested in the
general integration, specially if, in the cosmological context, creation probability
is desire computed. Recall that the last is written in terms of the potential
extracted from the WdW equation, namely,
\begin{equation}
{\cal P} \sim e^{-2\int ^{a_r} _{a_l} \sqrt{V}da}\,.
\end{equation}
In order to get some interesting results from the quantum approach, we shall consider
some special cases.

\subsection{Case \bf{A}}




\noindent If $E=0$ from Eq. (\ref{eq:moreenergy}) then $\Upsilon$ is just a constant given by
\begin{equation}
\frac{(k_2 F/12M^2 _{(2)})^2}{H^6} = (1 -\Upsilon)^2(\Upsilon - \frac{\Lambda}{6H^2})\,.
\end{equation}
The probability rate in this case is
\begin{equation}
{\cal P} \sim e^{\frac{4((\Upsilon -1) -\Lambda/3H^2)}{\Upsilon \Lambda} +
2(\Upsilon -1)H^2(\frac{6}{\Lambda})^2[1 -\frac{1}{X}\tan ^{-1} X]} ,
\label{eq:prob}
\end{equation}
where  $X^2 = (\frac{\Lambda}{6H^2})^2 \left(\Upsilon - \frac{\Lambda}{6H^2} \right)^{-1}$.
Now, if $k_2 F, \Lambda << H^2$ and, at first order the probality rate is
\begin{equation}
{\cal P} \sim e^{- \frac{4}{3H^2} + \frac{16k_2 F}{15H^5}}.
\label{eq:prob1}
\end{equation}
This means that it is more probable to create a universe when $k_2F > 0$ than $k_2F < 0$.
We will comment about it below.

Now, we would like calculate the probability nucleation using the instanton method.

The corresponding Euclidean action in de Sitter bulk can be found by complexifying
the temporal coordinate and keeping the field strength $F_{\mu \nu \rho \delta \gamma}$ fixed
\begin{equation}
S_{(E)}=  \int_m d^4 x \sqrt{-\gamma} \,\left(-\frac{M^2 _{(2)}}{2}{\cal R} + \rho_v \right) +
\frac{k_2}{4!} \int_m d^4 x \sqrt{-\gamma} A_{\mu \nu \rho \sigma }
\epsilon^{\mu \nu \rho \sigma } \,.
\label{eq:instanton}
\end{equation}
In Euclidean space we have now closed worldsheets that split the deSitter background spacetime
of radius $H_{dS} ^{-1}= (\Lambda/6)^{-1/2}$ in two regions. This is the basic geometry
of the instanton calculation.

Following \cite{Garriga2}, by using Stoke's theorem we
can transform (\ref{eq:instanton}) to an instanton action that involves a
volume of the spacetime enclosed by the brane
\begin{equation}
S_{(E)}=  \int_m d^4 x \sqrt{-\gamma} \,\left(-\frac{M^2 _{(2)}}{2}{\cal R} + \rho_v \right) -
k_2 F \int_v d^5 x \sqrt{-g}\,.
\end{equation}
For spherical worlsheets the former action is expressed through the radius $R_0$ of the brane
\begin{equation}
S_{(E)}= \left( \rho_v - \frac{12M^2 _{(4)}}{R^2} \right)  S_4 (R_0) - k_2 F V_4 (R_0) ,
\label{eq:Euclidean}
\end{equation}
where
\begin{equation}
S_{(4)} = \frac{8\pi^2}{3}R_0 ^4 ,
\end{equation}
is the surface of a worldsheet of radius $R_0$, and
\begin{equation}
V_{4} = \pi^2 H_{dS} ^{-5} \phi_0 - \frac{\pi^2 H_{dS}^{-4}}{R_0}
( 1 - R_0 H_{dS})^{1/2} ( 1 + \frac{2}{3} R_0 ),
\end{equation}
is the volume enclosed by the brane of radius $R_0$ and $ \sin(\phi_0) =
R_0 H_{{dS}}$. Extremizing (\ref{eq:Euclidean}) we find that the radius of
the Euclidean brane is a solution of
\begin{equation}
M^2_{(4)} H^3 \left( \Upsilon - \frac{\Lambda}{6H^2}\right)^{1/2}(1-\Upsilon) = \frac{k_2F}{12},
\end{equation}
where $\Upsilon \equiv H_{dS}^2 (R_0 H) ^{-2}$. The resulting Euclidean action is
\begin{equation}
S_{(E)} = -6\pi^2 M^2 _{(4)}\left\{\frac{4\left[(\Upsilon -1) -
\frac{\Lambda}{3H^2}\right]}{\Upsilon \Lambda} + 2(\Upsilon -1)
\left(\frac{6H}{\Lambda}\right)^2 \left[1 -\frac{1}{X}\tan ^{-1} X \right] \right\} ,
\end{equation}
and the nucleation probability ${\cal P} \sim e^{-S_{(E)}}$ is in agreement
with (\ref{eq:prob}) modulo a normalizing factor. We now go back to the
meaning of equation (\ref{eq:prob1}). The behavior of strength field
$F_{\mu \nu \rho \delta \gamma}$ is the key, when $k_2 >0$ the field decrease in
the inside region with respect to its original value and corresponds to screening
membrane discuss in \cite{Garriga2}. When $k_2 < 0$ correspond to antiscreening
membrane and the field increase its value, and as it is expected, is less probable
to produce such a Universe. This situation is resembled in phenomena of vacuum
decay, where ordinary transition from false to true vacuum corresponds to $k_2 >0$,
and the decay of true vacuum, by means of false vaccum bubbles, corresponds to $k_2 < 0$
and $k_2F$ represents the difference in energy density between the false and
true vacuum.

\subsection{Case \bf{B}}




\noindent We proceed to calculate an approximate expression for the
nucleation rate at first order,  when both $E$ and $F$ are small. The potential is
\begin{equation}
V(a) = 4a^2( 1 -H^2a^2 -EH - k_2FHa^4)
\end{equation}
and the nucleation probability is
\begin{equation}
{\cal P} \sim e^{-\frac{4}{3H^2} + EH^{-1} + \frac{16k_2F}{15H^5}}
\end{equation}
in complete agreement with (\ref{eq:prob1}) when $E$ vanishies.

\begin{figure}[h]
\epsfxsize=10cm
\centerline{\epsfbox{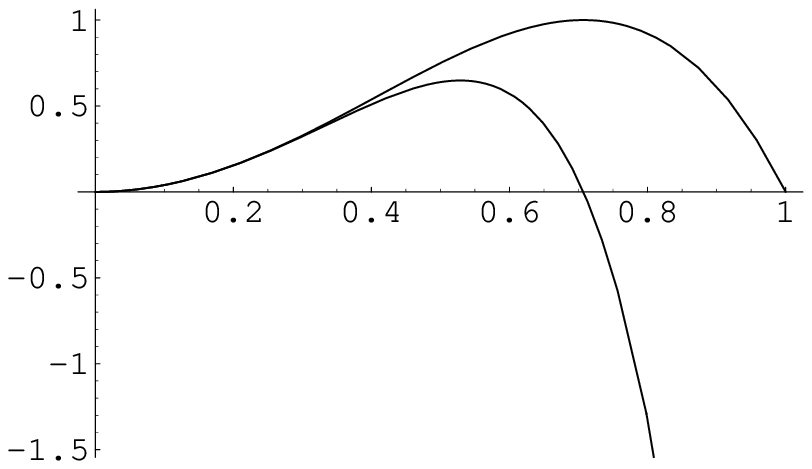}}
\caption{Potential for case A. In this case $E=0$ and $k= k_2 F$
taking the values: $k= 0$ (Einstein case) for the upper curve and $k\neq 0$
for the lower curve.}
\label{fig:1}   
\epsfxsize=10cm
\centerline{\epsfbox{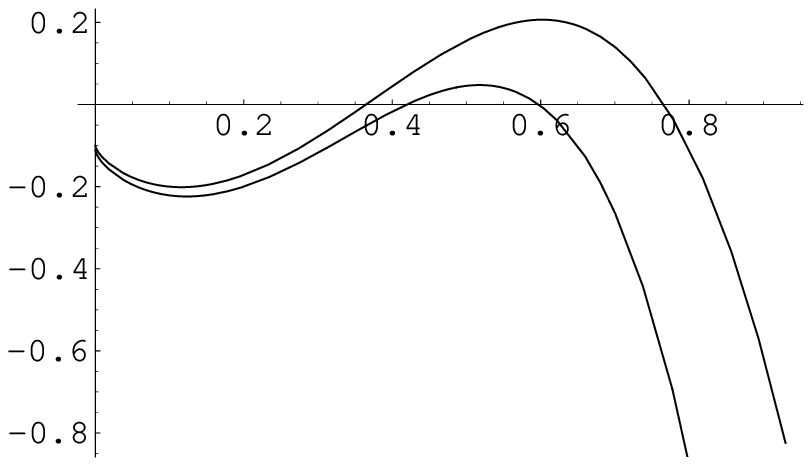}}
\caption{Potential for case B. In this case $E\neq 0$ and the background
is a de Sitter space. $k= 0$ for the upper curve and $k\neq 0$ for the lower curve. }
\label{fig:2}
\end{figure}

The potential for case A, is plotted in figure (\ref{fig:1}) and the corresponding
one for the case B is in figure (\ref{fig:2}). Using this kind of plots for the potential,
we can deduce that creation probability is enhance when the nucleation process take
place in  de Sitter background spacetime with small radius $H_{dS} ^{-1}$.

\section{Conclusions}

We have calculated the nucleation probability of brane world universes
induced by a totally antisymmetric tensor living in a dS fixed background
spacetime. This was done by means of canonical quantum approach where
the Wheeler-DeWitt equation was found. Besides, we found
for one specific case, the nucleation rate computing the corresponding
instanton. When the energy of the brane ${\cal{E}}=0$ in the bulk space
and the coupling constant of the brane $k_2$ with the antisymmetric
field is positive, the creation probability is enhanced with respect to
no interaction of the brane with the 4-form. For $k_2<0$ the nucleation
rate decresed as is expected. This situation is resembled in phenomena
of vacuum decay, where ordinary transition from false to true vacuum
corresponds to $k_2 >0$, and the decay of true vacuum by means of false
vaccum bubbles corresponds to $k_2 < 0$. Furthermore, $k_2F$ represents the
difference in energy density between the false and true vacuum.

For large expansion rate of the de Sitter bulk we observed an increase
nucleation rate. At this point we ask ourselves about possible brane
collisions, and what the most important factor in this issue is. The
branes will be driven apart by the exponential expansion of the bulk
reducing brane collision but at the same time, there is an increase in
nucleation rate. We expect now that the problem of old inflationary model
of the universe is an advantage: bubbles may not be produced fast enough,
to complete cover the bulk.

Once the brane universe was created it still could be hitting by stealth
branes \cite{Ruben-Alex}, that by means of constraining some parameters of the model reduce
the rate of brane collisions to an acceptable level.
We think that cosmological constraints can impose bounds on the values of $k_2F$
and with this value one could try to answer the question: Is our universe very
special?

\bigskip
{\bf Acknowledgements}

We benefited from Germ\'an Mandujano for assistance.
ER would like to thank C\'esar de la Cruz,
Ra\'ul Hern\'andez, Carlos Vargas and Alfredo Villegas for useful
discussions and encourage the paper.
We also thank to SNI-M\'exico for partial support.

\bigskip

\appendice{\\
\bf Embedding theory}

Consider a brane, $\Sigma$, of dimension $d$ whose worldsheet,
$m$ is an oriented timelike manifold living in a $N$-dimensional
arbitrary fixed background spacetime $M$ with metric $g_{\mu \nu}$.
For hamiltonian purposes, we shall foliate the worldsheet
$m$ in spacelike leaves $\Sigma_t$.

Taking advantage of the differential geometry for surfaces,
as well as novelty variational techniques developed in
\cite{Defo-DefoEdges,Jemal} we can write the Gauss-Weingarten
equations associated with the embedding of $\Sigma_t$ in $M$ ($x^\mu = X^\mu
(u^A)$), i.e., the gradients of the $\Sigma_t$ basis
$\{ \epsilon^\mu {}_A, \eta^\mu, n^\mu {}_i \}$.
These spacetime vectors can be decomposed with respect
to the adapted basis to $\Sigma_t$, as
\begin{eqnarray}
{\cal D}_A \epsilon^\mu {}_A &=& - \Gamma^\mu _{\alpha \beta}\,
\epsilon^\alpha {}_A \epsilon^\beta {}_B + k_{AB}\,\eta^\mu
- K_{AB} ^i \,n^\mu {}_i
\label{eq:GW1} \\
{\cal D}_A \eta^\mu &=& k_{AB} \,\epsilon^{\mu \,B}
- K_A {}^i \,n^\mu {}_i
\label{eq:GW2} \\
\tilde{{\cal D}}_A n^{\mu\,i} &=& K_{AB} ^i \,\epsilon^{\mu \,B}
- K_A {}^i \,\eta^\mu
\label{eq:GW3}
\end{eqnarray}
where $\Gamma^\alpha _{\beta \gamma}$ are the Christoffel
coefficients of the background manifold and,
$K_A {}^i$ is a piece of the generalized extrinsic twist
potential and both $k_{AB}$ and $K_{AB} ^i$ are the extrinsic
curvatures of $\Sigma_t$ associated with the normals
$\eta^\mu$ and $n^\mu {}_i$, respectively. ${\cal D}_A$ denotes
the covariant derivative adapted to $\Sigma_t$ and $\tilde{{\cal D}}_a$
is the covariant derivative that preserves invariance under rotations
of the normals $n^\mu {}_i$, i.e., $\tilde{{\cal D}}_A ^i = {\cal D}_A ^i
- \omega_A ^{ij} \, n_j$. In a similar way,
we can write the Gauss-Weingarten
equations associated with the embedding of $\Sigma_t$ in the
worldsheet $m$, ($x^a = X^a (u^A)$), i.e., the gradients
of the $\Sigma_t$ basis
$\{ \epsilon^a {}_A, \eta^a \}$.
These worldsheet vectors can be decomposed with respect
to the adapted basis to $\Sigma_t$, as
\begin{eqnarray}
\nabla_A \epsilon^a {}_B &=& \gamma_{AB}^C \,\epsilon^a {}_C
+ k_{AB}\,\eta^a
\label{eq:GW4}
\\
\nabla_A \eta^a &=& k_{AB} \,\epsilon^{a \,B} \,,
\label{eq:GW5}
\end{eqnarray}
where $\nabla_A$ is the gradient along the tangent basis, i.e.,
$ \nabla_A = \epsilon^a {}_A \nabla_a$, where $\nabla_a$ is the covariant
derivative compatible with $\gamma_{ab}$.

The time vector field, written in terms of the adapted basis of
a leaf $\Sigma_t$, is given by
\begin{equation}
t^\mu = \dot{X}^\mu = N \eta^\mu + N^A\,\epsilon^\mu {}_A \,,
\label{eq:timevector1}
\end{equation}
which represents the flow of time throughout spacetime.
Note that we are able to rewrite the previous time
deformation vector as follows
\begin{eqnarray}
\nabla X^\mu &:=& t^a \nabla_a X^\mu - N^A {\cal D}_A X^\mu
\nonumber \\
&=& N\,\eta^\mu \,,
\label{eq:timevector2}
\end{eqnarray}
where, taking into account the well known notation,
$\nabla_a$ denotes the covariant derivative compatible
with $\gamma_{ab}$
\quad ($\mu, \nu = 0,1,2,\ldots,N-1$;
$a,b= 0,1,\ldots,d$ and $A,B= 1,2,\ldots, d$).
Furthermore, from (\ref{eq:timevector1}) note
that the following relations hold:
$$
N= -g_{\mu \nu} \eta^\mu \dot{X}^\mu \quad \quad
{\mbox{and}} \quad \quad
N^A = g_{\mu \nu} h^{AB}\epsilon^\mu {}_A \dot{X}^\nu .
$$

\bigskip

\appendice{\\
\bf $\Psi$ \quad Matrix}

In this appendix we write the full matrix $\Psi$ for our embedding
(\ref{eq:embedding}). Taking into account the Eq.
(\ref{eq:tensor-matrix}) as well as Eq. (\ref{eq:lambda}) we have
\begin{equation}
(\Psi)^{\mu \nu}=
\left( \begin{smallmatrix}
-\frac{1}{2 a^2 A_{\pm}}\left[ 6 + \Lambda_b a^2 + \frac{6\dot{a}^2}{(-\Delta)}
\right] & 0 &
0 & 0\\
0 & \frac{A_{\pm}}{2a^2}\left[ 6 + \Lambda_b a^2 + \frac{6\dot{a}^2}{(-\Delta)} +
12 A_{\pm} \right]& 0 & 0  \\
0 & 0& \frac{1}{2a^4}\left[ 6 + \Lambda_b a^2 + \frac{6\dot{a}^2}{(-\Delta)} \right]
& 0 \\
0 & 0 & 0 &
M_{2\times 2}
\end{smallmatrix} \right) .
\end{equation}
The previous matrix, in the cosmic gauge, reduces to a more manageable form
\begin{equation}
(\Psi)^{\mu \nu}=
\left( \begin{smallmatrix}
3H^2 A_{\pm} ^{-1}(1 - \Upsilon ) & 0 & 0 & 0 \\
0 & 3 A_{\pm} a^{-2} [- H^2 a^2 (1 - \Upsilon ) + 2 A_{\pm} ]& 0 & 0  \\
0 & 0& - 3 a^{-2} H^2 (1 - \Upsilon ) & 0 \\
0 & 0 & 0 &  N_{2\times 2}
\end{smallmatrix} \right),
\label{eq:Psicosmicgauge}
\end{equation}
where $M_{2 \times 2}$ and $N_{2\times 2}$ denote $2 \times 2$ diagonal
matrices.

\vfill\break
\bibliographystyle{unsrt}





\end{document}